\documentclass[12pt]{article}

\usepackage{amsmath,amsthm,amssymb,latexsym %,mathrsfs
}
\usepackage{graphicx,color,array,fullpage,fancyhdr,cancel}
\usepackage{hyperref}
\usepackage{url}
\usepackage{tikz}
\usepackage{enumitem}
\usepackage{soul}
\usepackage{cite}
\usepackage{bbm}
\usepackage[margin=1.02in]{geometry}

\newtheorem{theorem}{Theorem}[section]

\theoremstyle{definition}

\theoremstyle{remark}

\numberwithin{equation}{section}

\begin{document}

\title{Fractal AC circuits and propagating waves on fractals}

\author{Eric Akkermans\footnote{
Department of Physics, Technion - Israel Institute of Technology, 32000 Haifa, Israel
}
\\
Joe P. Chen\footnote{ 
Department of Mathematics, Colgate University, Hamilton, NY 13346, USA}
\\
Gerald Dunne\footnote{
Department of   
Physics, University of Connecticut, Storrs, CT 06269, USA
}
\\
Luke G. Rogers\footnote{ 
Department of Mathematics, University of Connecticut, Storrs, CT 06269, USA
}
\\
Alexander Teplyaev$^\S$}

\date{\today 
%\thanks{
%This paper was first completed in July 2015 and posted on the arXiv as \href{https://arxiv.org/abs/1507.05682}{\texttt{arXiv:1507.05682}}. 
%}
}

\maketitle

%\thanks{Research supported in part by NSF grants DMS-1106982 and DMS-1262929}

\begin{abstract}
We extend Feynman's analysis of the  infinite ladder AC circuit to   fractal AC circuits. We show that
 the characteristic impedances can have positive real part even though all the individual impedances inside the circuit  are purely imaginary. This provides a physical setting for analyzing  wave propagation of signals on fractals, by analogy with the Telegrapher's Equation, and generalizes the real resistance metric on a fractal, which provides a measure of distance on a fractal,  to complex impedances.
\end{abstract}

\section{Introduction}\label{sec-intro}

\begin{figure}[h]
 \center\includegraphics[scale=.5]{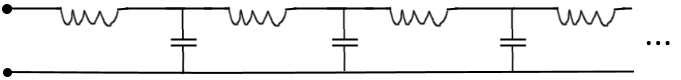}
 \caption{The Feynman infinite ladder $LC$ circuit \cite{F}.}
 \label{fig-ladder}
 \end{figure}

It is well known that fractal-like structures have novel spectral and response properties, both classically and quantum mechanically. However, the concept of wave propagation on fractals has been a long-standing puzzle, since even the notion of velocity is unclear on such singular geometrical structures. Here we propose a new approach to this problem, by considering a fractal circuit comprised of inductors and capacitors. Part of our motivation comes from  the familiar textbook example of the infinite ladder circuit, described for example  in Feynman's lectures \cite{F}: see Figure~\ref{fig-ladder}. This is an important example relating AC circuits to wave propagation, as these ladders are directly related to the wave equation known as the Telegrapher's Equation (see, for example, \cite{EvansPDE}). In more general terms, the introduction of complex impedances in fractal circuits leads to a physical picture of energy flow in a fractal. In fractal circuits consisting only of resistors, the effective resistance, derived from Kirchhoff's laws, gives a concrete physical realization of the static distribution of energy in the circuit. This idea is a key element  in the theory of energy forms on fractals, and provides a definition of a `metric' on a fractal, leading also to a bridge between spectral properties and geometry. 
%Our goal is to extend some of these ideas to the complex domain in order to describe the time-dependent flow of energy ina  fractal. 
Algebraically, this analysis can be extended to include inductance and capacitance, using the Kirchhoff rules combined with the same geometric properties of the fractal. Surprisingly, the results are quite different, and describe situations in which signals propagate though the fractal in a manner very different from conventional circuits.
This simple idea should have profound consequences for the description of time-dependent phenomena on fractals, as well as leading to AC circuits with novel energy transport properties.
%Fractals composed solely of resistors have been well-studied and the effective resistance on such fractals is a key concept in defining distance metrics, energy forms and spectral properties of Laplacians, on fractals. 
 
The conventional infinite ladder circuit is shown in Figure~\ref{fig-ladder}. It is 
 made of infinite sequences of inductances $L$ and capacitors $C$, and 
iterating the circuit elements leads to
 the characteristic complex impedance 
 at frequency $\omega$: 
 $$Z=\frac i{2C}\left(\omega LC+\sqrt{\omega^2L^2C^2-4LC}\right) $$ 
 if  $\omega^2LC>4$, but 
 $$Z=\frac1{2C}\left(i\omega LC+\sqrt{4LC-\omega^2L^2C^2}\right) $$ 
 if  $\omega^2LC<4$. 
 Note that in the latter case the characteristic impedance has a positive real part even though all elements in the circuit have purely imaginary impedances. 
This circuit behaves as a low pass filter---admitting signals with low frequencies and damping out signals with high frequencies---because  the  propagation factor $\alpha$ has $|\alpha|<1$ for $\omega^2LC>4$, and  $|\alpha|=1$ for $\omega^2LC<4$. 
 In this review paper, we show how to extend Feynman's analysis to   fractal networks.
Further computational 
analysis is developed in \cite{AC-power}, and spatial intermittency 
and singularly of the energy measure at infinity is proved in \cite{AR}. We refer the interested reader to these works for more details.
In particular it is explained in \cite{yoon, AC-power}    how to take the appropriate limits and choosing the branch of the complex square root. 
 
 Our analysis, although simple in nature, relies on the ideas of some recent papers on physics on fractals \cite{ADT09,ADT10,ABDTV12,Akk,Dunne12} 
 which are motivated by mesoscopic physics, \cite{AM,DalNegro}, some recent connections between gravity and fractals \cite{Ambjorn:2005db,Reuter:2012xf}, and early physics papers on fractals  \cite{AO,DABK,GAA,RT,sgmagnetic} which highlighted the unusual physical properties of fractal structures. For physics applications 
such as \cite{minati2018high,kempkes2018design}, it is important to note that even though some of these results have been proven mathematically for infinitely-iterated mathematical fractals, many of the unusual physical features are visible even for physical finitely-iterated  structures which technically speaking are not mathematical fractals. 
% The singularity of e
 Energy measures 
 in related real circuits are analyzed in 
 \cite{BST99,hino}.  Our analysis of the Hanoi-type graphs (discussed below) is motivated by related analytic 
  \cite{hanoi,2012hanoi} and algebraic \cite{NT,KSW} features. 
Furthermore, the Hanoi-type graphs play an important role in \cite{MRT,FST}, 
which is closely related to the physical renormalization group discussion in  \cite{englert}. 
 In more general terms, 
 our work  relies on the understanding of 
 energy in fractal resistance networks, as developed in  
 \cite{Str06,Ngasket,fractalina,twists2,HMT,BBKT,grad,Tcjm,PT,OSCT,hinz2013dirac,hinz2015metric,kelleher2018differential,2012hanoi,hinz2016magnetic,brzoska2017spectra,hinz2017viscous,falconer1999non} 
and related spectral analysis 
\cite{FSh,T,MT1,MT2,Str03,ST12,vib1,IPRRS,Ro,3Ngasket,twists1,SteiT,SteiJMP,chen2016singularly,post2006spectral}.  

In \cite{AC-power} we prove 
that any physically relevant self-similar AC circuit on the Sierpinski gasket 
is a constant complex multiple of a purely real self-similar   circuit. This simple fact has been an obstacle to progress in this direction. 
Therefore one needs to modify the standard construction in order to obtain more interesting properties. 
We present the two representative examples 
of such modifications in the following sections. 
The first is based on the standard graph approximation of the Sierpinski gasket, which is referred to as the \emph{Feynman-Sierpinski ladder circuit} \cite{AC-power,AR}.
The second is based on the Hanoi graph approximation of the Sierpinski gasket.

\section{Fractal modified Sierpinski circuits}\label{sec-SG}

\begin{figure}[htb]
\center \includegraphics[scale=.5]{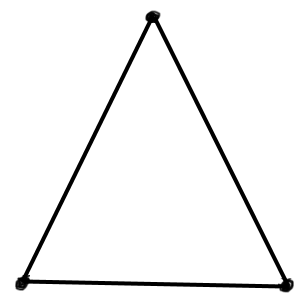} \qquad \includegraphics[scale=.5]{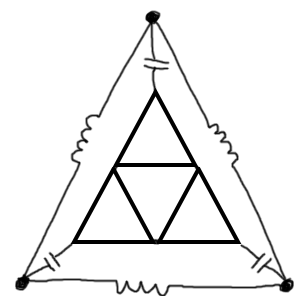} \qquad
\includegraphics[scale=.5]{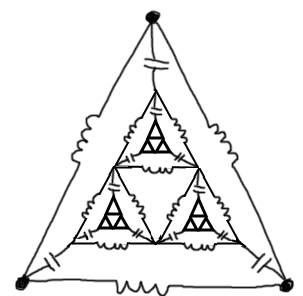}
\caption{The Feynman-Sierpinski ladder circuit.}\label{fig-SG}
\end{figure}

Our first non-trivial fractal AC circuit is constructed by a substitution procedure, the first two steps of which are shown in Figure~\ref{fig-SG}.  In each case the upward-pointing triangles with dark outlines are replaced by the circuit in the central diagram.  This takes the triangle at the left to the central circuit, and the central circuit to that at the right.  The substitutions are repeated infinitely many times to obtain the final circuit, which is  made of purely imaginary impedances $i\omega L$ and $1/(i\omega C)$.   
Our  main idea is that the modified Sierpinski gasket (SG) circuit contains 3 copies of itself, which allows for the fractal version of Feynman's analysis.

\begin{theorem}
\label{thm:Z}
%If we d
Denote the impedance of all sides of the original triangular circuit  at the left in  Figure~\ref{fig-SG} by $Z$ and set it equal to the corresponding impedance of the limiting circuit. Then
%The infinite 
 % modified SG circuit, obtained by substituting each solid  triangle by the circuit in the middle of Figure~\ref{fig-SG}, 
%has the characteristic impedance of the triangle with side impedance 
\begin{equation} \label{e01-mSG}
Z=\frac1{10\omega C}\left(2i\omega^2LC
+9i+ \sqrt{144\omega^2LC-4(\omega^2LC)^2-81}\right)
\end{equation}
provided
\begin{equation} \label{e01-mSG-f}9(4-\sqrt{15})<2\omega^2LC<9(4+\sqrt{15}).
\end{equation}  
In particular, this circuit is a filter for this frequency range.  
Outside the range \eqref{e01-mSG-f} we have 
\begin{equation} \label{e01-mSG2}
Z=\frac i{10\omega C}\left(2\omega^2LC
+9- \sqrt{4(\omega^2LC)^2+81-144\omega^2LC}\right) \text{\qquad if \  $9(4-\sqrt{15})>2\omega^2LC$}
\end{equation}
\begin{equation} \label{e01-mSG3}
Z=\frac i{10\omega C}\left(2\omega^2LC
+9+ \sqrt{4(\omega^2LC)^2+81-144\omega^2LC}\right) \text{\qquad if \  $9(4+\sqrt{15})<2\omega^2LC$}
\end{equation}
 and so $Z$ is a purely imaginary impedance under these conditions. 
 \end{theorem}
 
As a sanity check, the $+$ (resp.\@ $-$) sign in front of the square root can be 
verified by considering the special case of zero frequency limit $\omega\to0$ (resp.\@ infinite frequency limit $\omega\to\infty$).

\begin{proof}[Proof of Theorem \ref{thm:Z}]
%To obtain these results, 
Observe that the Sierpinski 
gasket has the resistance scaling factor 5/3. This remains true for complex impedances because it depends only on the symmetries in the geometric structure of the gasket. Therefore the  three solid triangles of impedances $Z$ are equivalent to one triangle with side impedance $5Z/3$. We can then do a delta-Y transform to obtain an Y-shaped circuit where each leg has impedance 
$5Z/9+1/(i\omega C)$, followed by a Y-delta transform to obtain
%\begin{equation} \label{e01-mSG-1}
$\dfrac{1}{Z}= \dfrac1{i\omega L}+\dfrac1{5Z/3+3/(i\omega C)},
$
which implies~\eqref{e01-mSG}.
%\end{equation} 
    \end{proof}

An interesting additional fact proved in~\cite[Theorem 3.1]{AC-power} is that the characteristic impedances of the sequence of finite approximations of the Feynman-Sierpinski ladder circuit fail to converge to the impedance $Z$ in~\eqref{e01-mSG}, but that introducing a small positive resistance $\epsilon$ in series with each of  the capacitors and inductors gives a sequence of scale $N$ approximating circuits for which the impedances  $Z_{n,\epsilon}$ converge as $N\to\infty$. Moreover for the regularized limit $\lim_{\epsilon\to0+}\lim_{N\to\infty}Z_{N,\epsilon}=Z$. 

\begin{figure}[h]
\center \includegraphics[scale=1.11]{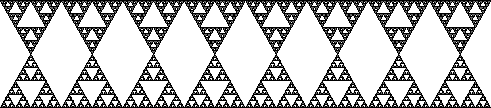}
 \caption{The   infinite ladder fractafold, \cite{chen2015spectral,Str03,ST12} .}\label{fig-ladder-f}
 \end{figure}

 For a physical configuration analogous to a two-sided AC ladder circuit, we can consider the infinite ladder fractafold shown in Figure~\ref{fig-ladder-f}. Replacing the Sierpinski elements with the AC fractal SG-type circuits constructed by the process in Figure~\ref{fig-SG}, we can construct an infinite fractafold ladder with transmission properties very different from those of the ladders in Figure~\ref{fig-ladder}. 
Understanding the connection between (static) spectral analysis on fractafolds  \cite{chen2015spectral,Str03,ST12}
and dynamical AC behavior is an interesting open question.

\section{Fractal Hanoi circuits}\label{sec-hanoi}

Another class of fractal AC circuits can be constructed as follows.
We begin with a bilaterally symmetric Y-shaped circuit, \begin{figure}[htb]
\center \includegraphics[scale=.2]{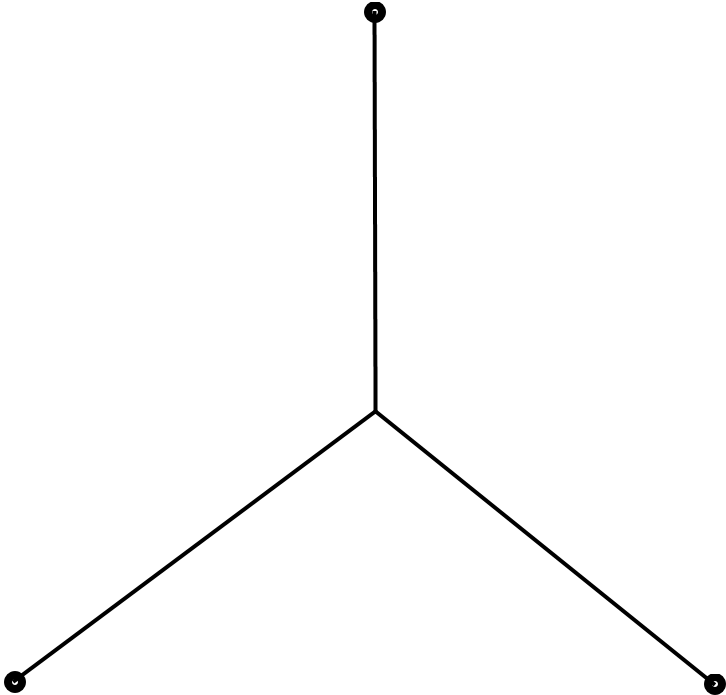} \qquad \includegraphics[scale=.2]{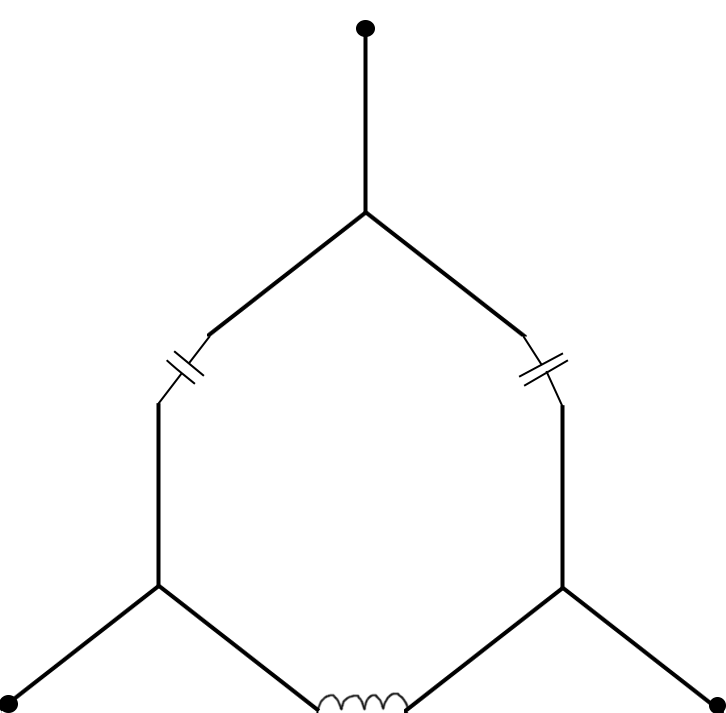} \qquad
\includegraphics[scale=.2]{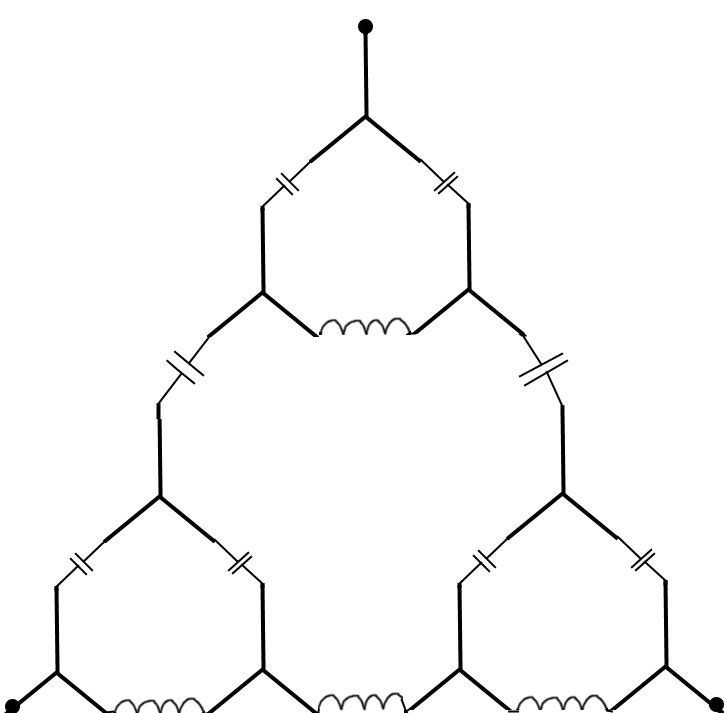}
\caption{A fractal Hanoi-type circuit construction.}\label{fig-Hanoi1}
\end{figure} 
defined by the vertical impedance $Z_v$, and left and right impedances $Z_{\text{L}}=Z_{\text{R}}$, to obtain a weakly self-similar AC fractal circuit   such that 
this original circuit 
is equivalent to the Hanoi-type circuit shown in Figure~\ref{fig-Hanoi1}.  The   idea is, again, that the Hanoi circuit contains rescaled copies on the original Y circuit, which allows for the fractal  analysis to proceed. The main difference is that when  we go from scale to scale we reduce all impedances by 
a factor $r$. Thus at the $n$th iteration of the construction, each of the $3^n$ Y-shaped pieces have the vertical  impedance $r^nZ_v$ and left and right impedances $r^nZ_{\text{L}}$. The inductances and capacitors are $r^{k-1}L$  and $r^{1-k}C$, where $k$ is the scale when this particular element first appeared in the construction. 
We can   compute effective impedance top-to-bottom and left-to-right to obtain the following two equations:
\begin{equation}\label{e01-yh}
Z_v+Z_{\text{L}}/2=rZ_v+(rZ_v+2rZ_{\text{L}}+1/(i\omega C))/2
\end{equation}
and
\begin{equation}\label{e02-yh}
2Z_{\text{L}}=2rZ_{\text{L}}+\left(\dfrac1{2rZ_{\text{L}}+i\omega L}+\dfrac1{2rZ_v+2rZ_{\text{L}}+2/(i\omega C)}\right)^{-1}.
\end{equation}
From \eqref{e01-yh} we obtain 
\begin{equation}\label{e01-yh-}
(2-3r)Z_v+(1-2r)Z_{\text{L}}=\frac1{i\omega C}
\end{equation}
which implies  the following special case: 
\begin{equation}\label{e02-yh-r1/2}
\text{if \ $r=\dfrac12$ \ then \ $Z_v=\dfrac2{i\omega C} $ \ and \ $Z_{\text{L}}=Z_{\text{\tiny\sc{r}}}=2\sqrt{L/C}$}. 
\end{equation}
In this case \eqref{e02-yh}  can be simplified to  $
\dfrac1{Z_{\text{L}}}= {\dfrac1{Z_{\text{L}}+i\omega L}+\dfrac1{Z_{\text{L}}+4/(i\omega C)}} 
$. Equation \eqref{e02-yh-r1/2} is  remarkable not only because  it has a purely real impedance for an infinite circuit of imaginary impedances, but  also   this real impedance is independent of $\omega$. Similar behavior in a different simpler circuit was obtained numerically  in \cite{dragon}. 
%Another interesting special case is: 
%\begin{equation}\label{e02-yh-r0}
%\text{if \ $r=0$ \ then \  $Z_v=\frac{1}{i\omega C(2-\omega^2LC)} $ \ and \ $Z_{\text{L}}=Z_{\text{\tiny\sc{r}}}=\frac{i\omega L}{2-\omega^2LC}$}. 
%\end{equation}

In general, we have 
characteristic impedances with positive real part when
$0<r<\frac35$ and 
\begin{align}\label{e-range}
\gamma(r) - \sqrt{[\gamma(r)]^2 -1} < 2LC\omega^2 < \gamma(r) + \sqrt{[\gamma(r)]^2 -1}.
\end{align}
where
\begin{align}\label{e-gamma}
\gamma(r) = 1+\frac{r(3-5r)}{(2r-1)^2}.
\end{align}
Further details are given in \cite[Lemma 4.1 \& Theorem 4.2]{AC-power}, including the determination of the aforementioned low-pass filter condition from the quadratic equation %s 
\begin{align}
%\label{e-Zv}
%&r(5r-3)i\omega CZ_v^2
%+(2r-1)(2-\omega^2LC)Z_v+i\omega L=0
%\\
\label{e-Zell}
&r(5r-3)i\omega CZ_{\text{L}}^2
+(2r-1)(2-\omega^2LC)Z_{\text{L}}+i\omega L=0.
\end{align}
% leading to convergence properties for 
 %$0\leqslant r<\dfrac35$ and non-convergence for 
 %$r\geqslant\dfrac35$.
In particular, if  $r\in [\frac{3}{5},1)$  then there are solutions for $Z_{\text{L}}$ and $Z_{\text{R}}$, but the filer condition fails because $Z_{\text{R}}$ is purely imaginary for all frequencies.
Curiously, the critical $r$-value $\frac{3}{5}$ is well known to be the renormalization factor for fully symmetric resistances on the Hanoi graph to converge to a limiting energy form on the fractal.
We do not have an intuitive explanation for why this occurs in the AC circuit context.

 \section{Conclusions}
 We have shown how to construct nontrivial AC circuits with fractal geometrical structure which have novel transmission properties. This provides a novel experimental and theoretical framework for investigating the propagation of waves on fractal structures. It also generalizes the concept of a ``spatial'' resistance-metric to a ``space-time'' setting.
 
\subsection*{Acknowledgments}
Research supported in part by 
NSF grants DMS-1106982, DMS-1613025, DMS-1262929  and DMS-1659643,
DOE grant ER41989, DOE grant DE-SC0010339, 
Simons Foundation Grant No.\@ 523544,
and Israel Science Foundation
Grant No.\@ 1232/13. 
The authors are very grateful 
to 
Patricia Alonso-Ruiz, 
Loren Anderson, 
Ulysses Andrews, 
Edith  Aromando, 
Antoni Brzoska, 
Aubrey  Coffey, 
Hannah Davis, 
Michael Dworken, 
Lee Fisher, 
Madeline  Hansalik, 
Stephen Loew
and
Daniel Kelleher 
for helpful discussions. 

\subsection*{Contact details}
 
  \url{http://physics.technion.ac.il/~eric/}  
   \href{mailto:eric@physics.technion.ac.il}{eric@physics.technion.ac.il}
 \\
 \url{http://math.colgate.edu/~jpchen/} 
 \href{mailto:jpchen@colgate.edu}{jpchen@colgate.edu}
 \\
 \url{http://dunne.physics.uconn.edu}
 \href{mailto:dunne@phys.uconn.edu}{dunne@phys.uconn.edu}
 \\
 \url{http://www.math.uconn.edu/~rogers/}
 \href{mailto:rogers@math.uconn.edu}{rogers@math.uconn.edu}
 \\
 \url{http://homepages.uconn.edu/teplyaev/}
 \href{mailto:teplyaev@math.uconn.edu}{teplyaev@math.uconn.edu}

 %\urladdr{\url{http://mathreu.uconn.edu/}}

 %\urladdr{\url{http://homepages.uconn.edu/fractals/fractals/}}
 
\bibliographystyle{ieeetr}
\bibliography{AC}

\end{document}